\documentclass[twocolumn,showpacs,preprintnumbers,amssymb]{revtex4}

\begin{document}
\title{Thermal and Electrical Currents in Nanoscale Electronic Interferometers}
\author{Sam Young Cho and Ross H. McKenzie}
\affiliation{Department of Physics, University of Queensland,
             Brisbane 4072, Australia 
}
\date{\today}

\begin{abstract}
 We theoretically study thermal transport 
 in an electronic interferometer 
 comprising a parallel circuit of two quantum dots, 
 each of which has a  tunable single electronic state which
 are connected to two leads at different temperature. 
 As a result of 
 quantum interference, 
 the heat current through one of the dots
 is in the opposite direction to the temperature gradient.
 An excess heat current flows through the other dot.
 Although locally, heat flows from cold to hot,
 globally the second law of thermodynamics
 is not violated because
 the entropy production associated with
 heat transfer through the whole device is still positive.
 The temperature gradient also induces
 a circulating electrical current, which makes the interferometer
 magnetically polarized.
 
\end{abstract}
%
%
%
%
%
\maketitle
 
 {\it Introduction.}
 Manipulation of quantum coherence and interference 
 in a controllable manner is of special interest
 in nanoscale electron devices \cite{review}. 
 The coherence of
 resonant electron tunneling through a quantum dot (QD)
 has been demonstrated by using Aharonov-Bohm interference \cite{Yacobi95}. 
 Moreover, such interference effects have enabled
 the realization of a phase sensitive probe of 
 the anomalous transmission phase \cite{Schuster97},
 dephasing effects \cite{Buks98}, and 
 many-body correlation effects \cite{Wiel00} 
 in quantum coherent transport through a QD. 
 Very recently,
 a quantum interferometer based on two QDs 
 has been fabricated and 
 control of coherent electron
 transport by varying gate voltages of each dot \cite{Holleitner00}
 has been demonstrated.
 In such a double dot interferometer,
 theoretical studies have focused on the subjects
 of resonant tunneling~\cite{Kubala02}, cotunneling~\cite{Akera93}, 
 many-body correlation effect~\cite{Izumida97}, 
 magnetic polarization current~\cite{Cho03},
 and two-electron entanglement in the context of quantum 
 communication \cite{Loss00}.
 Also, there has been considerable interest
 in thermal transport through nanoscale devices \cite{Sivan,Beenakker,Guttman,
 Moskalets2,Appleyard,Schwab,Andreev,Matveev,Kim02,Humphrey02}
 and possible ``violation'' of 
 the second law of thermodynamics for small colloidal
 systems over short time scales
 \cite{Wang}, small quantum systems \cite{Nie}, and nanoscale electric circuits 
 \cite{Nie2}.

\begin{figure}
\vspace*{5.00cm}
\includegraphics{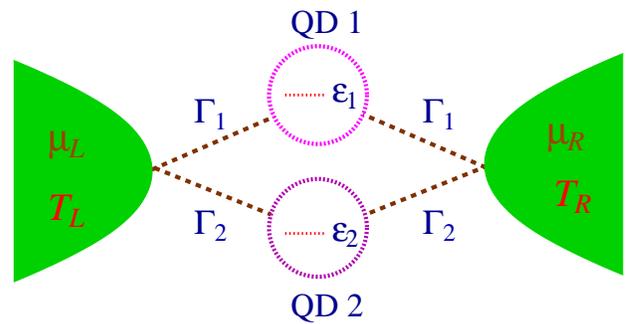}
\caption{(color online) A quantum interferometer based on two quantum dots. 
  Both dots are tunnel-coupled to the left and right leads.
  The tunneling amplitudes between the dots and the leads 
  are denoted by $\Gamma_1$ and $\Gamma_2$.
  Each lead is described by an equilibrium Fermi-Dirac 
  distribution with temperature $T_L$ and $T_R$ and
  electrochemical potential $\mu_L$ and $\mu_R$.
  The energy level position in each dot is measured 
  as $\varepsilon_1$ and $\varepsilon_2$ relative to the Fermi energy
  in the leads.
  }
\label{fig1}
\end{figure}

 In this paper,
 we consider 
 the thermal transport induced by a temperature gradient 
 across a double dot interferometer
 (see Fig. \ref{fig1}). 
 The thermal transport could be manipulated
 in a controlled manner such as
 varying gate potentials.
 In the interferometer, electric current conservation does not require
 that the total current through the interferometer should be
 greater than the local current through each electron path.
 The quantum interference of tunneling electrons
 results in a circulating electric current which 
 can make the magnetic states of the device be up-, non-, or down-polarized.
 It was recently shown that the magnetic polarization current
 exists at a finite bias between the leads \cite{Cho03}.
 In this study,
 the temperature difference between the leads
 can give rise 
 to a circulating electric current without an applied bias.
 Furthermore, it is found that due to quantum interference
 the heat current flows in the opposite direction
 to the temperature drop through one dot
 while the excess heat current flows through the other dot.
 The behaviors of the local heat currents show
 the existence of a circulating heat current. 
 We discuss the second law of thermodynamics associated with
 the two unique thermal transport processes in the interferometer.
 
%
{\it Model.}
 We start with a general model Hamiltonian
\begin{equation}
  H  = \sum_{\alpha} H_\alpha \{ c^\dagger_{k\sigma}; c_{k\sigma} \}
     + \sum_{\alpha,j} H^T_{\alpha,j}
     + \sum_{j} H^D_{j} \{ d^\dagger_{j\sigma}; d_{j\sigma} \},
\label{Hamiltonian}
\end{equation}
 where $\alpha(=L,R)$ and $j(=1,2)$ are the indices of leads and dots.
 The Hamiltonians $H_\alpha$, $H^D_j$, and 
 $$H^{T}_{\alpha,j}=\sum_{k\sigma \in \alpha;j}
 \left[ 
 V^{\alpha}_{k,j} c^\dagger_{k\sigma} d_{j\sigma} + {\rm h.c.} 
 \right],$$
 respectively
 represent the leads, the interferometer, and tunneling between
 the leads and the dots.
 $c_{k\sigma}$ and $d_{j\sigma}$ are the annihilation operators 
 with spin $\sigma$ 
 for electrons in the leads and the dots.
 
%
 The energy (number of electrons) flowing into the interferometer 
 is defined as the rate of change in the energy (number of electrons)
 in the lead $\alpha$ \cite{Mahan}:
 $I^E_\alpha =-{d \langle H_\alpha \rangle / dt}
 = (i/ \hbar) \langle [ H_\alpha, H ] \rangle$ and 
 $I_\alpha =-{d \langle N_\alpha \rangle / dt}
 = (i/ \hbar) \langle [ N_\alpha, H ] \rangle$, 
 where $H_\alpha=\sum_{k\sigma\in \alpha} \varepsilon_k 
  c^\dagger_{k\sigma} c_{k\sigma}$ and
  $N_\alpha=\sum_{k\sigma\in \alpha} 
  c^\dagger_{k\sigma} c_{k\sigma}$.
 The heat current flowing into the interferometer from the lead $\alpha$
 is defined by
 \begin{equation}
  I^Q_\alpha = I^E_\alpha - \mu_\alpha I_\alpha,
 \end{equation}
 where $\mu_\alpha$ is the electrochemical potential in the lead $\alpha$.
 Using the Keldysh Green function 
 $G^<_{k\sigma,j\sigma}(t-t') \equiv i\langle d^\dagger_{j\sigma}(t') 
                               c_{k\sigma}(t) \rangle$
 which involves electron 
 operators for the leads and for each dot,
 one writes the heat currents as 
 \begin{eqnarray}
   I^Q_\alpha 
   \!\! &=& \!\!
   - \!\! \sum_{k\sigma\in \alpha}\!\! \int\!\! \frac{d\varepsilon }{2\pi\hbar}
     (\varepsilon_k -\mu_\alpha)
    \left[ V_{k,1} G^<_{k\sigma,1\sigma}(\varepsilon) + {\rm h.c.} \right]
   \nonumber \\ && 
       \!\! - \!\!
       \sum_{k\sigma \in \alpha}\!\! \int \!\!\frac{d\varepsilon}{2\pi\hbar}
     (\varepsilon_k -\mu_\alpha)
    \left[ V_{k,2} G^<_{k\sigma,2\sigma}(\varepsilon) + {\rm h.c.} \right] \, .
 \label{current}
 \end{eqnarray}
 The first (second) line of Eq.~(\ref{current}) describes heat
 transfer from the left lead to QD 1 (QD 2) or vice versa.
 Then each heat transfer 
 can be defined as a local heat current through each dot, $I^Q_{\alpha,j}$.
 Thus the total heat current is the sum of the local heat currents 
 through each dot,
 $I^Q_\alpha = I^Q_{\alpha,1} + I^Q_{\alpha,2}$.
 The heat current is written in terms of
 the electron Green functions, $G^<$. 
 The heat transfer from the lead to one of the dots
 is then accompanied by electron dynamics including 
 a complex trajectory through the entire interferometer,
 as well as a direct tunneling to the dot.

 The Green functions, $G^<$, can be expressed in terms of 
 the dot Green functions defined by
 $G_{jj',\sigma}^{\rm r}(t-t')
 =-i\theta (t-t')\langle \{ d_{j'\sigma}(t),d_{j\sigma}^\dagger(t')\}\rangle$
 and
 $G_{jj',\sigma}^{<}(t-t')
 =i\langle d_{j'\sigma}(t)d_{j\sigma}^\dagger(t')\rangle$ \cite{Meir92}.
 As a consequence, a general expression of the heat current 
 through a nanoscale electron interferometer is given by
 \begin{eqnarray}
  I^Q_\alpha
       &=&
       i \sum_{jj'\sigma}  \int
         \frac{d\varepsilon}{2\pi\hbar}
       (\varepsilon-\mu_\alpha)
         \Gamma_{jj',\alpha}(\varepsilon)
   \left[ 
         G^<_{jj',\sigma}(\varepsilon)
      \right. \nonumber \\ && \hspace*{1.4cm} \left.
       + f_\alpha(\varepsilon)
        \left(  G^r_{jj',\sigma}(\varepsilon)
              - G^a_{jj',\sigma}(\varepsilon) \right)
   \right],
 \end{eqnarray}
 where the tunnel couplings between the leads
 and the dots are denoted by 
 $\Gamma_{jj',\alpha}=2\pi {\cal N} V^\alpha_{j} V^{\alpha *}_{j'}$
 with the density of states of the lead ${\cal N}$.
 The Fermi-Dirac distribution functions of the leads
 are
 $f_\alpha (\varepsilon)=f(\varepsilon-\mu_\alpha)$,
 where
 the chemical potentials are
 $\mu_L=-\mu_R=eV/2$ with applied bias $V$ between the two leads.
 For $i \neq j$, the terms of the current describe 
 the interference between the electron waves through two dots.
 In the absence of one dot, i.e., 
 $V_{k,1}=0$ or $V_{k,2}=0$, only the current 
 through the other dot exists and
 any interference resulting from the existence of the one dot
 disappears.
 Therefore, the expression of the heat current is reduced to
 the heat current formula in a single dot electronic device. 

 Our interferometer has the two electron pathways which
 are allowed for electron transport from one lead to the 
 other via two dots not being coupled to each other directly.
 Electron tunneling through the dots are
 manipulated by varying the gate voltages.
 The dots makes it possible to  
 control a coherent electron passing through
 the two electron pathways in the interferometer where 
 it is required to satisfy current conservation
 at the leads. The current conservation
 gives rise to a circulating current on
 a closed path through the dots and the leads \cite{Cho03}.
 To clarify the origin of the circulating electric/heat currents 
 in the interferometer
 the intradot electron-electron interaction
 is not taken into account in this study.
 Then the level spacing in each dot is larger than
 the applied bias and temperature
 because electron transport through a single level in the dots.
 Although intradot Coulomb interactions are considered
 in the Coulomb blockade regime, 
 the resonant transport could be well explained
 in the Hatree-Fock mean-field level where 
 the energy level of the dots can be described by
 a simple shift of the interaction parameter.
 In fact, we focus on studying the interference effects
 that are present for near resonant transport
 and employ the resonant level model to describe the dots;
 $ 
  H^D_j = \sum_{j\sigma} \varepsilon_j d^\dagger_{j\sigma} d_{j \sigma},
 $ 
 where $\varepsilon_1$ and $\varepsilon_2$ are the level energy in each dot,
 measured, relative to the Fermi energy of the leads.


 With the Keldysh technique
 for nonlinear current through the system,
 the local heat currents through each dot at the lead $\alpha$
 are given by
 \cite{Meir92,Cho99},
\begin{equation}
 I^Q_{\alpha,j} = \sum_\sigma \int \frac{d \varepsilon}{2\pi\hbar} \, 
   (\varepsilon -\mu_\alpha)
    (f_\alpha(\varepsilon)-f_{\alpha'}(\varepsilon)) {\cal T}_j(\varepsilon),
 \label{heat}
\end{equation}
 and similarly the local electric currents are obtained as
\begin{equation}
   I_{\alpha,j} = 
   e \sum_\sigma \int \frac{d \varepsilon}{2\pi\hbar} \, 
   (f_\alpha(\varepsilon)-f_{\alpha'}(\varepsilon)) {\cal T}_j(\varepsilon),
 \label{electric}
\end{equation}
 where the local transmission spectral functions are defined by
   ${\cal T}_j(\varepsilon)= 
     \left\{ \mbox{\boldmath $\Gamma$}^L 
             {\bf G}^{\rm r}_\sigma (\varepsilon)
             \mbox{\boldmath $\Gamma$}^R 
             {\bf G}^{\rm a}_\sigma (\varepsilon) \right\}_{jj} $
 which is the $j$-th diagonal component of the matrix transmission
 spectral function.
 ${\bf G}^{\rm r}_\sigma (\varepsilon)$ 
 is the matrix dot Green function defined in time space as 
 $G_{jj',\sigma}^{\rm r}(t-t')
 =-i\theta (t-t')\langle \{ d_{j'\sigma}(t),d_{j\sigma}^\dagger(t')\}\rangle$.
 The matrix coupling to the leads is described by
 ${\bf \Gamma}^L={\bf \Gamma}^R=
 \left(\begin{array}{cc} \Gamma_1 & \sqrt{\Gamma_1 \Gamma_2 } 
  \\ \sqrt{\Gamma_1 \Gamma_2 } & \Gamma_2 \end{array} \right)$.
 The symmetric tunnel-coupling between the dots and the leads 
 will be assumed to be independent of energy.
 The matrix Green function of the dots is 
 \begin{equation}
 \label{Green's function}
   {\bf G}_\sigma^{\rm r}(\varepsilon) =
        \left( \begin{array}{cc} 
        \varepsilon-\varepsilon_1 + i \Gamma_1
        & i \sqrt{\Gamma_1 \Gamma_2}  \\
          i \sqrt{\Gamma_1 \Gamma_2} 
        & \varepsilon -\varepsilon_2+i \Gamma_2
        \end{array} \right)^{-1}.
 \end{equation}
 From the relation,
 ${\bf G}_\sigma^{\rm a}(\varepsilon)
      =[{\bf G}_\sigma^{\rm r}(\varepsilon)]^\dagger$,
 the advanced Green function can be obtained.
 Accordingly, the local transmission spectral functions 
 in terms of the total transmission function are given by
\begin{eqnarray}
  {\cal T}_1(\varepsilon) &=&
  {\Gamma_1 (\varepsilon -\varepsilon_2) 
   \over
  \Gamma_2(\varepsilon \! - \! \varepsilon_1)
     +\Gamma_1(\varepsilon \! -\! \varepsilon_2)} {\cal T}(\varepsilon) , \\
  {\cal T}_2(\varepsilon) &=&
  {\Gamma_2 (\varepsilon -\varepsilon_1) 
   \over
  \Gamma_2(\varepsilon \! - \! \varepsilon_1)
     +\Gamma_1(\varepsilon \! -\! \varepsilon_2)} {\cal T}(\varepsilon) .
\end{eqnarray}
%
 The total current is the sum of current through each dot,
 $I^Q=I^Q_1+I^Q_2$, which is just the current conservation.
 This leads to the total transmission spectral function as
 ${\cal T}(\varepsilon)={\cal T}_1(\varepsilon)+{\cal T}_2(\varepsilon)$,
\begin{equation}
  {\cal T}(\varepsilon) \! = \!\!
  {
  \left[\Gamma_2(\varepsilon\! -\! \varepsilon_1)
     \!  +\!\Gamma_1\!(\varepsilon\! -\!\varepsilon_2)\right]^2 
   \over
  (\varepsilon \!-\! \varepsilon_1)^2 
        (\varepsilon\!-\!\varepsilon_2)^2
   \! + \!
  \left[\Gamma_2(\varepsilon \! - \! \varepsilon_1)
     +\Gamma_1(\varepsilon \! -\! \varepsilon_2)\right]^2 } .
\end{equation}

 In the linear response regime, the transport electric/heat currents
 are expanded
 up to the linear terms of $\Delta T = T_L-T_R$ and $\Delta V = V_L - V_R$.
 The electric currents and the heat currents are related 
 to the chemical potential difference, $\Delta V$, 
 and the temperature difference, $\Delta T$,
 by the thermo-electric coefficients $L_{mm'}$ :
 \begin{equation}
  \left( \begin{array}{c} I \\ I^Q \end{array} \right)
  =
  \left( \begin{array}{cc} L_{11} & L_{12} \\ L_{21} & L_{22} 
          \end{array} \right)
  \left( \begin{array}{c} \Delta V \\ \Delta T \end{array} \right).
 \label{linearI}
 \end{equation}
 Similarly, with the local thermo-electric coefficients $L^{(j)}_{mm'}$
 the local electric/heat currents in the linear
 response regime can be written as 
 \begin{equation}
  \left( \begin{array}{c} I_{j} \\ I^Q_{j} \end{array} \right)
  =
  \left( \begin{array}{cc} L^{(j)}_{11} & L^{(j)}_{12} \\ 
                           L^{(j)}_{21} & L^{(j)}_{22} 
          \end{array} \right)
  \left( \begin{array}{c} \Delta V \\ \Delta T \end{array} \right).
 \end{equation}
 The thermo-electric coefficients associated with the local current
 through each dot
 are expressed as
 $L^{(j)}_{11} = e^2 {\cal L}^{(j)}_0$, $L^{(j)}_{21}=TL^{(j)}_{12}=-e{\cal L}^{(j)}_1$,
 and $L^{(j)}_{22}={\cal L}^{(j)}_2/T$, where 
 the integrals are defined as
$
 {\cal L}^{(j)}_n (T) 
  = \frac{2}{h} \int d\varepsilon 
     \left(-\frac{\partial f}{\partial \varepsilon} \right)
      \varepsilon^n {\cal T}_j(\varepsilon)$.
 According to the current conservations, the thermo-electric
 coefficients have the relations:
 $L_{mm'}=L^{(1)}_{mm'}+L^{(2)}_{mm'}$.

\begin{figure}
\vspace*{6cm}
\includegraphics{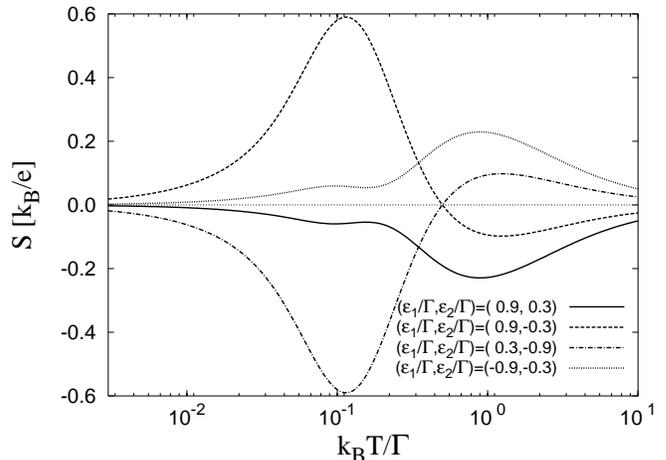}
\caption{Thermopower of the interferometer as a function of temperature
         for different level positions of the dots, 
         $(\varepsilon_1/\Gamma,\varepsilon_2/\Gamma)$. 
         The tunneling amplitudes are taken as $\Gamma=\Gamma_1=\Gamma_2$.
         For both energy levels of dots being above (below) the
         Fermi energy, the sign of the thermopower is negative (positive).
         The charge and heat are carried mainly through 
         the electron (hole) channels.
         When the energy level of one dot is lying below the Fermi
         energy and that of the other is lying above the Fermi energy,
         the sign of the thermopower is changed as 
         the temperature varies. Therefore,
         the main propagating channels for 
         charge and heat determine the sign of the thermopower.
         The magnitude of the thermopower is of the order of 
         $k_B/e = 86.17 \mu {\rm V}/K$.
         }
 \label{fig2}
\end{figure}

 {\it Thermopower.}
 The thermopower of the interferometer
 can be found by measuring the induced voltage drop
 across the interferometer when the temperature difference
 between two leads is applied.
 For zero electric transport current, $I=0$, 
 the thermopower is defined by the relation
  \begin{equation}
  S = - \lim_{\Delta T \rightarrow 0} 
             \frac{\Delta V}{\Delta T} \bigg|_{I=0}
     =  \frac{L_{12}}{L_{11}}.
  \end{equation}
 In terms of the defined integrals, 
 one can rewrite the thermopower,
 $ S= -(k_B/e)({\cal L}_0/k_BT {\cal L}_1) $
 with the constant $k_B/e\simeq 86.17 \mu V/K$.
 In Fig. \ref{fig2}, 
 the characteristics of the thermopower are shown to be 
 dependent on the energy level positions of the dots. 
 The sign of the thermopower can indicate
 the main channel in transporting charge and heat.
 When more transmission spectral weight lies in the electron channel
 then in the hole channel, 
 the charge and heat are carried by mainly electron channels.
 In this case the sign of the thermopower is negative.
 In the opposite case, 
 since charge and heat transport through the hole channels is 
 predominant,
 the thermopower is positive.
 If the same amount of electron and heat are carried by each
 electron and hole channel, 
 the sign of the electric/heat current is the same/opposite
 for electron and hole channel.
 This results in the thermopower being zero.
 As shown in Fig. \ref{fig2},
 when the energy level of one dot is lying below the Fermi energy
 and that of the other dot is lying above the Fermi energy,
 the sign of the thermopower is changed as temperature increases.

\begin{figure}
\vspace*{6cm}
\includegraphics{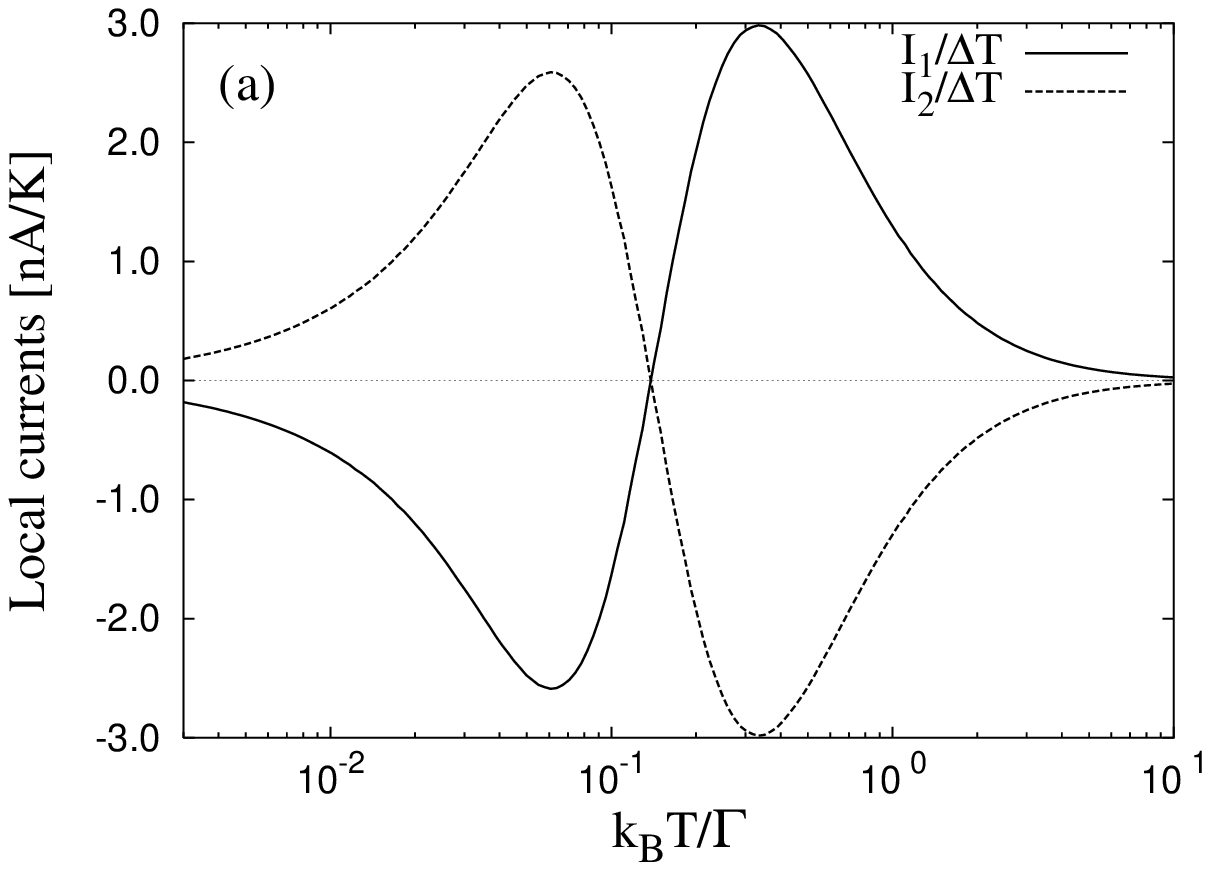}
\vspace*{6cm}
\includegraphics{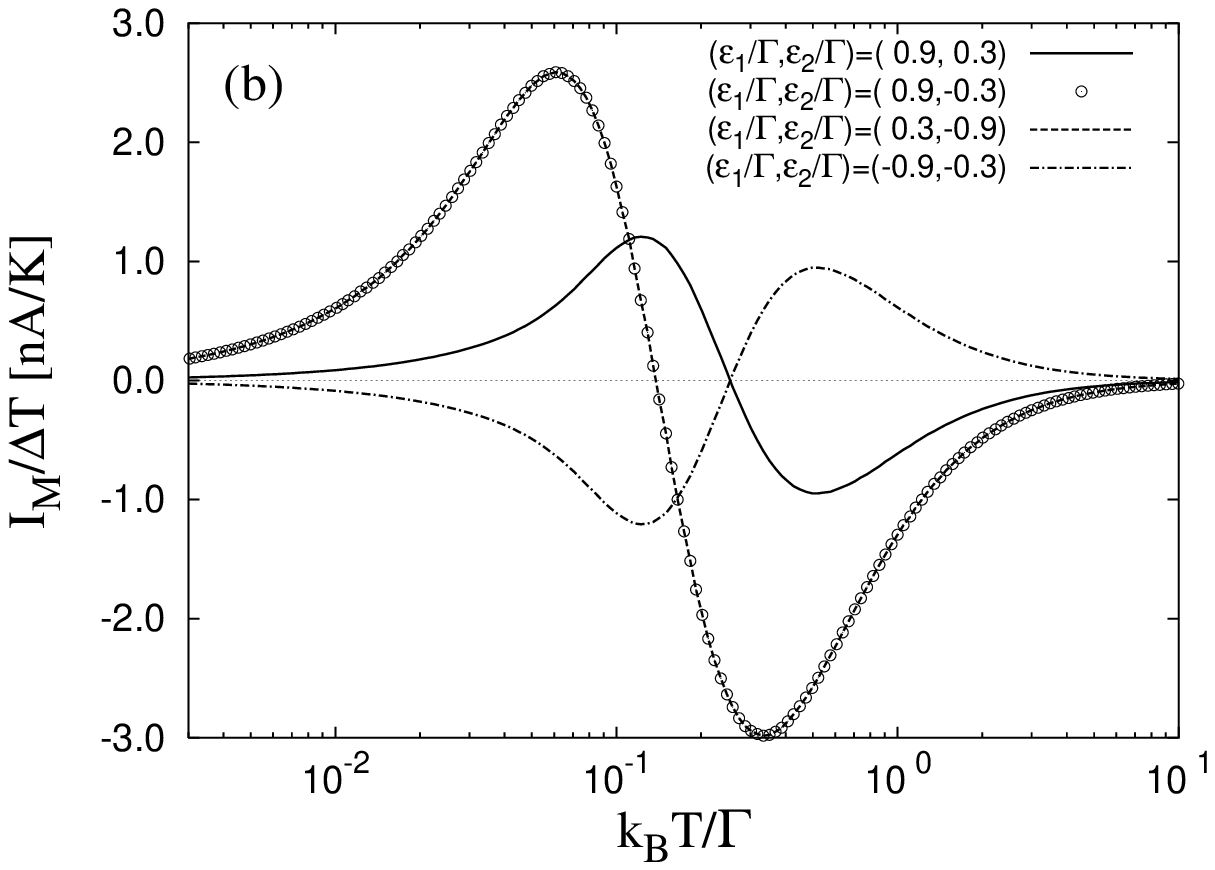}
\caption{(a) Local electric currents induced by a temperature gradient
         as a function of temperature. 
         The dot energy levels are taken as
         $(\varepsilon_1/\Gamma,\varepsilon_2/\Gamma)=(-0.9,0.3)$
         for the tunneling amplitudes, $\Gamma=\Gamma_1=\Gamma_2$.
         In contrast to the thermopower (compare Fig. \ref{fig2}), 
         the level positions taken in 
         this plot do not affect the physics of the local currents
         but only change its amplitude and sign.
         Even when the total electric transport current is zero $(I=0)$,
         the non-vanishing local electric currents indicates
         the existence of the circulating electric current
         which makes the interferometer magnetically polarized.
         The magnetic polarization current is then defined
         as the local current: $I_M \equiv I_1 = -I_2$ for $I=0$.
         (b)
         The ratio of the magnetic polarization currents to the temperature
         gradient, $\Delta T$, as a function of temperature.
         The magnitude of the magnetic polarization currents
          is of the order of ${\rm nA}$ for a temperature gradient of order
          $1 {\rm K}$.
         }
 \label{fig3}
\end{figure}

\begin{figure}
 \vspace*{5.5cm}
 \includegraphics{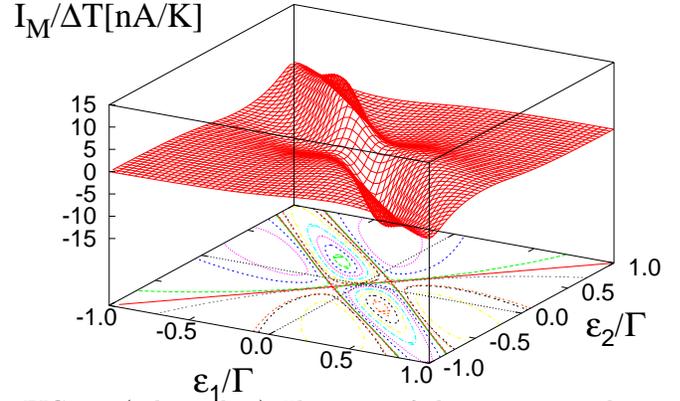}
 \caption{(color online) The ratio of the magnetic polarization current 
          $I_M \equiv I_1=-I_2$ 
          to the temperature gradient $\Delta T$
          for $I=0$ as a function of the energy level
          positions of each dot $(\varepsilon_1/\Gamma, \varepsilon_2/\Gamma)$
          for $k_B T = 5.0 \times 10^{-2} \Gamma$.
          The tunneling amplitudes are taken as $\Gamma=\Gamma_1=\Gamma_2$.
          Note that by varying the energy level of each dot the interferometer
          has a different magnetic state according to the direction of the
          magnetic polarization current.}
\label{fig4}
 \end{figure}

 {\it Local electric currents and magnetic polarization currents.}
 The requirement, that the electric transport current is zero,
 for the thermopower 
 implies that the local electric currents are required 
 to ensure $I_1 = -I_2$.
 If these local currents exist for $I=0$, 
 the local electric currents should circulate on the closed path 
 through the leads and the dots.
 Then the interferometer can
 be magnetized by the circulating electric currents.
 One can define the circulating current as 
 a magnetic polarization current \cite{Cho03}, $I_M \equiv I_1=-I_2$ for $I=0$.
 From Eq. (\ref{linearI}), the magnetic polarization current
 is then expressed as
 \begin{equation}
 I_M = {\cal K} \, \Delta T,
 \end{equation}
 where
 ${\cal K}= - S L^{(1)}_{11} + L^{(1)}_{12}
     = S L^{(2)}_{11} - L^{(2)}_{12}$. 
 This shows that the magnetic polarization current exists
 even when the electric transport current is zero
 because this magnetic polarization current is induced 
 by the temperature gradient
 between the leads due to the quantum interference.
 Figure \ref{fig3}(a) shows that the total sum of the local currents
 is always zero because the local electric currents
 are flowing along the opposite direction to each other, which
 implies the existence of the magnetic polarization current.
 It should be noted that the direction of the magnetic polarization current 
 is reversed as the temperature increases. 
 If one can define the local thermopower as $S_j = L^{(j)}_{12} /L^{(j)}_{11}$,
 the magnetic polarization current
 vanishes at a specific temperature, $T_0$, satisfying $S(T_0) =-S_j(T_0)$.
 It is also shown in Fig. \ref{fig4} (b) that,
 by manipulating the gate voltages of each dot,
 the magnetic polarization current can be controlled.
 This implies that changing the energy level positions of each dot,
 one can magnetize the interferometer by the magnetic polarization
 current induced by the temperature gradient as
 up-, non-, and down-polarized.
 In contrast to the measurement of the thermopower by means of 
 electron transport,
 to observe the magnetic polarization current,
 one can measure a magnetic field produced by the magnetic polarization 
 current by using a superconducting quantum interference device (SQUID). 
 Recent measurements of a persistent current by using a SQUID \cite{Rabaud01}
 show that
 magnitudes of the magnetic polarization current 
 of the order of ${\rm nA}$ 
 should be experimentally observable
 for a temperature gradient of order $0.1 \sim 1 {\rm K}$.

\begin{figure}
 \vspace*{5.5cm}
 \includegraphics{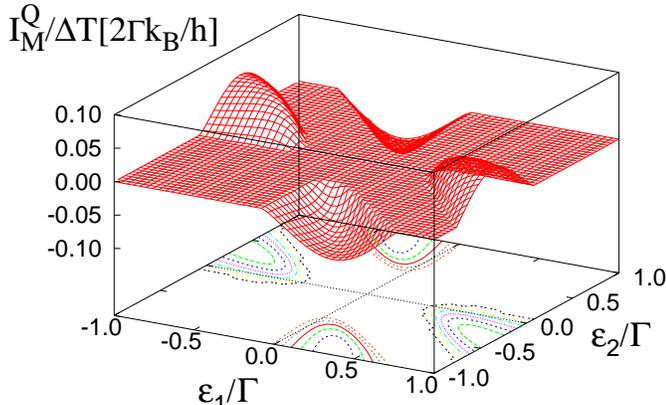}
 \caption{(color online) The ratio of the circulating heat current 
          to the temperature gradient $\Delta T$  
          as a function of the energy level
          positions of each dot $(\varepsilon_1/\Gamma, \varepsilon_2/\Gamma)$
          for $I=0$.
          The parameters are taken as  $k_B T = 5.0 \times 10^{-2} \Gamma$
          and $\Gamma=\Gamma_1=\Gamma_2$.
          The existence of the circulating heat current indicates
          that
          while the local heat current through one dot
          can be greater than the total transport 
          heat current through the interferometer
          due to the quantum interference,
          the local heat current through the other dot
          flows against the temperature gradient.
          The current conservation requires 
          that the local excess heat current circulate
          on the closed path in the interferometer.
          }
\label{fig5}
 \end{figure}

 {\it Heat currents.}
 The condition of an open circuit ($I=0$) to find the thermopower
 can apply for a circulating heat current in the interferometer. 
 Under the condition, $I=0$, the local heat currents are rewritten as
 \begin{equation}
  I^Q_j = \kappa_j \Delta T,
 \end{equation}
 where $\kappa_j = -S L^{(j)}_{21} + L^{(j)}_{22}$.
 Due to the quantum interference, the local heat currents
 can be greater than the total heat current through the interferometer
 at a given energy level positions of the dots.
 If $I^Q_1 > I^Q$, there exists an excess heat current
 through QD 1. 
 One can define the excess current as $I^Q_{exe} = I^Q_1 - I^Q$.
 From the relation $I^Q=I^Q_1+I^Q_2$, the local heat current through
 the QD 2 should be $I^Q_2 =-I^Q_{exe}$.
 The negative sign of the local heat current through the QD 2 implies
 that the heat current conservation requires a heat current flowing
 through the QD 2 against the temperature gradient.
 According to the second law of the thermodynamics, 
 the entropy production defined by $I^S \equiv I^Q/T > 0$
 should be greater than zero during thermoelectric process \cite{Guttman}.
 If one can define a local entropy production as $I^S_j \equiv I^Q_j/T$,
 the local heat current flowing against the temperature gradient
 means that $I^S_2 < 0$.
 However, for heat transfer through the entire interferometer,
 the heat current conservation should make
 the second law of the thermodynamics preserved
 in heat transport through the entire interferometer. 
 As a result, the excess heat transport through the QD 1 
 is compensated with the local heat current flowing against 
 the temperature gradient through the QD 2. 
 Like the magnetic polarization current,
 those thermoelectric processes imply 
 that there appears a circulating heat current
 on the closed path between the dots and the leads
 in order to satisfy the second law of the thermodynamics.
 Therefore, for $I^Q_1 > I^Q$,  
 the excess current can be defined as a circulating heat current, 
 $I^Q_M \equiv I^Q_{exe}=-I^Q_2$. 
 Similarly, for $I^Q < I^Q_2$, the circulating heat current is determined.
 In Fig. \ref{fig5}, we display 
 the circulating heat current as a function of 
 the energy level positions of each dot from the numerical calculation.
 It is shown that the interference between the electron and hole
 channels produces the circulating heat current.

{\it Summary.} 
 We have investigated thermal transport
 in nanoscale interferometers.
 The expression of the heat current for the interferometer
 has been derived based on
 the nonequilibrium Green's function technique.
 Controllable electronic states in two dots
 make it possible to manipulate the quantum interference
 which causes a heat current in the opposite direction
 to the temperature drop through one dot
 and an excess heat current through the other dot.
 The circulating electric current induced by
 a temperature gradient across the entire interferometer
 is sufficiently large that 
 it should be experimentally observable.

{\it Note added.}
After completion of this work, we become aware of some related 
work by Moskalets concerning a temperature-induced circulating electric current 
in a one-dimensional ballistic ring \cite{Moskalets1}.

{\it Acknowledgments.}
This work was supported by the University of Queensland and
the Australian Research Council.


\end{document}